\documentstyle[aps,prb,multicol,graphicx]{revtex}

\draft

\begin{document}

\title{A Novel Theory for High Temperature Superconductors considering
Inhomogeneous Charge Distributions}

\author{E. V. L. de Mello, E. S. Caixeiro and J. L. Gonzal\'ez }

\address{Departamento de F\'{\i}sica,
Universidade Federal Fluminense, av. Litor\^ania s/n, Niter\'oi, R.J.,
24210-340, Brazil}
\date{Received \today }
\maketitle

\begin{abstract}

We present a percolation theory for the high-$T_c$ oxides
pseudogap and $T_c$ dependence on the hole level.
The doping dependent inhomogeneous charge structure 
is modeled by a distribution
which may represent the stripe morphology and yield a 
spatial distribution of local $T_c(r)$. 
The temperature  onset of spatial dependent superconducting 
gap is identified with the 
vanishing of the pseudogap temperature $T^*$.
The transition to a superconducting state corresponds to the
percolation threshold among regions of different $T_c$. 
%The onset of vanishing gap  is identified as the onset 
%of the doping dependent  pseudogap at the temperature $T^*$. 
As a paradigm we use a  Hubbard Hamiltonian 
with a mean field  approximation to yield a doping and 
temperature dependent superconducting d-wave gap. 
We show here that this new approach reproduces the phase diagram,
explains and gives new insights on several experimental features
of high-$T_c$ oxides.
\end{abstract}

\begin{multicols}{2}
\section{Introduction}
 High-$T_c$ oxides has been discovered fifteen years ago\cite{BM} but 
many of their important properties remains not well understood.  
Among these, the pseudogap phenomenon, that is, a discrete structure
of the energy spectrum above $T_c$, identified by
several different experiments\cite{TS}, has its nature not yet
been clarified. Such open problem has attracted a lot of experimental 
and theoretical effort because it is general belief that its solution
is related to the understanding of the superconducting fundamental 
interaction. 
%The existence 
%of such gap structure in the metallic phase above the superconducting 
%critical temperature $T_c$ and with a constant value at the transition
%is very unusual from the predictions of the conventional theory of
%superconductivity.

The evidence of such energy gap above
the  superconducting phase 
%and remains constant at this transition, 
is clearly demonstrated by tunneling\cite{Retal98,Setal99} 
and angle-resolved photoemission spectroscopy\cite{Shen95,Ding96}
experiments. In the resistivity measurements  
its presence is seen by  a decrease in the linear
behavior below $T^*$\cite{Oda00,Takagi92} and in the specific heat as
a suppression in the linear coefficient $\gamma$ of 
the temperature\cite{Loram97}. 
%These and many other important 
%experiments like nuclear
%magnetic resonance, magnetic neutron scattering and electron Raman
%scattering and their connection with the pseudogap have been discussed in 
%detail in Ref.\cite{TS}.

%As a consequence of this great experimental effort,  the existence 
%of a pseudogap structure above  $T_c$ became universally accepted and 
%The idea that the pseudogap  and the inhomogeneous charge distribution 
%must be dealed by a theory of high-$T_c$ oxides has recently
%attracted considerable attention. 
There is also mounting experimental evidence  that the
hole doped inhomogeneity into the $CuO_2$ planes,
common to  all cuprates,
is directly related to the pseudogap phenomenon. 
In a given family, the  underdoped compounds near the doping onset
of superconductivity have the 
more inhomogeneous charge distributions and the 
larger $T^*$. As the doping level increases, 
the samples become more homogeneous while $T^*$ 
decreases\cite{Billinge00,Buzin00,Pan}. For overdoped compounds
$T^*$ disappears or becomes equal to $T_c$.
The inhomogeneities were long supposed to be important in
the studies of high temperature superconductors\cite{Egami96} but 
only after
the discovered of the spin-charge stripes\cite{Tranquada},
they have become matter of systematical studies.
In the spin-charge stripes scenario, regions of the plane are heavily
doped (the stripes) and other regions are underdoped and fill the space
between the charge-rich stripes. 
%Stripes phases occur due to the competition between the 
%antiferromagnetic interaction among magnetic ions  and Coulomb interaction
%between the charges, both of which favor localization and, on the
%other hand, the zero-point of the doped holes, which favor 
%delocalization of the charges.
%Experimentally, they are more easily
%detected in insulating materials, where they are static, but there
%is evidence of fluctuating strip correlations in metallic and 
%superconducting compounds\cite{EKT}.

%In this article we introduce a model to explain 
%the  phenomenology of high $T_c$ oxides like the pseudogap, the
%inhomogeneous charge distribution and the superconducting transition.
%There are two facts that support the connection between the pseudogap 
%and the charge inhomogeneities: first,
%the value of the pseudogap has its  maximum
%for low doping compounds 
%for which  $T^*$ is well above $T_c$ and decreases rapidly
%with increasing doping and vanishes for sufficiently overdoped samples. 
%Secondly, experiments strongly indicates a highly                               inhomogeneous distribution of charge
%for underdoped compounds which gradually
%become more homogeneous with increasing doping\cite{Billinge00,Buzin00,Pan}
%exactly where $T^*$ decreases. 

Recently, magnetic excitations as the vortex-like Nernst effect have
been reported above $T_c$\cite{Xu} and a local Meissner state, which
usually appears only in the superconducting phase,
has been seen  as a precursor to superconductivity\cite{IYS}. Such
inhomogeneous diamagnetic domains develop
near $T^*$ and grow continuously  as $T$ decreases towards $T_c$.
Near $T_c$, the domains appear to percolate, according Fig.3 from
Iguchi et al\cite{IYS}.

Based on all these experimental findings, about
the pseudogap phenomenon, local charge inhomogeneities and
a non percolative local Meissner state between $T^*$ and  $T_c$,
we propose  a new scenario to
explain the  high $T_c$ superconductors phenomenology:
due to the doping dependent charge inhomogeneities in a given compound 
with average charge density $\rho_m$,
there is a distribution of local clusters with
spatial dependent charge density $\rho(r)$, each with its 
superconducting transition 
temperature $T_c(r)$. $T^*$ is the maximum of all $T_c(r)$.
As the temperature falls below  $T^*$, 
some clusters become superconducting,
but they are surrounded by metallic and/or antiferromagnetic 
insulating domains and, consequently, the whole system is not
a superconductor.
The number of superconducting clusters 
increases as the temperature decreases, so the superconducting regions grow
and, eventually, at a temperature $T_c$, they percolate and
become able to hold a macroscopic dissipationless current. 
Exactly as the Meissner state domains shown in Fig.3 of Iguchi et al\cite{IYS}.
Similar ideas were discussed by   
Ovchinnikov at al\cite{OWK}. They were  concerned mainly  with the
microscopic mechanism which leads to a distribution of $T_c(r)$
and its effect on the density of state.
%Stripe phases
%and lattice effects were also discussed with a two component 
%Hubbard model by Bussmann-Holder et al\cite{Annette,Egami}.

In order to show that these ideas are able to make quantitative predictions
and reproduce the high-$T_c$ oxides phase diagram, 
we have performed calculations on a Hubbard model 
%with a chemical 
%potential which controls the charge concentration 
and a gap equation is obtained within mean field approach. 
%To represent
%the doping dependent charge inhomogeneities, we used a 
%Poisson-Gaussian distribution although the main results are independent
%of the distribution details and can be derived with other distribution
%forms. 
%The details are given  in the following sections.
%
\section{The Charge Distribution}

%The presence of microscopic charge inhomogeneities distribution 
%in the $CuO_2$ planes, possibly in a striped configuration, is well
%documented by several experiments\cite{Tranquada,Bianconi,Billinge00,Buzin00}.
%It also implies in strong modifications for the local structure as the
%local $Cu-O$ bond length changes its size with the average copper 
%charge\cite{Buzin00,Egami}.

The consequence  of the microscopic charge inhomogeneities distribution
in the $CuO_2$ planes, possibly in a striped configuration,
is the existence of 
%domain walls between the 
two phases which are spontaneously created in the $Cu-O_2$ planes; 
regions which are heavily doped or hole-rich form 
the stripes and others regions which are hole-poor and 
are created between the charge-rich stripes. The exactly  
form of these charge distributions is not known and it is presently 
matter of research\cite{Billinge00,Buzin00,Egami}. 
%In this article, 
%in order to approximately  represent these two phase,  
%we introduce a model distribution.
We have chosen  a distribution capable to reproduce
the experimental observations, and for this purpose, 
we use a combination of a Poisson and a 
Gaussian distribution. 
For a given compound  with an average charge density $\rho_m$, the 
hole distribution is function of the local hole density $\rho$,
$P(\rho;\rho_m)$ divided in two branches.
The low density branch represents the  hole-poor or non-conducting regions and
the high density one represents the  hole-rich or metallic regions. 
As concerns the  superconductivity, only the properties of the hole
rich branch are important since  the current flows through the 
metallic region. Such distribution may be given by: 
%We have also
%performed calculations with low  branches which ended
%at $\rho=0.05$, which represents the hole-poor stripes, 
%but here we present a more symmetrical 
%model which both branches vanishing at $\rho_m$, given  by;

\begin{eqnarray}
P(\rho) &=& \pm (\rho-\rho_c)exp(-(\rho-\rho_c)^2/2(\sigma_{\pm})^2)/ \nonumber 
      \\ &&           (\sigma^2)(2-exp(-(0.05)^2/2(\sigma_{\pm})^2)) 
\end{eqnarray}

The plus sign is for the hole-rich ($\rho_c \approx \rho_m$) for
$\rho_m \le \rho$,
the minus to the hole-poor branch ($\rho_c= 0.05$) for 
$\rho \le 0.05$ and $P(\rho)=0$ for $0.05\le \rho \le \rho_m$. 
The half-width $ \sigma$ is related with the 
degree of inhomogeneities and decreases with  the hole density 
%for a given family and
%approaches the optimal doping value and for the 
%homogeneous overdoped compounds 
in order to represent current observations\cite{Billinge00,Buzin00,Egami}.

\begin{figure}
\includegraphics[width=8cm]{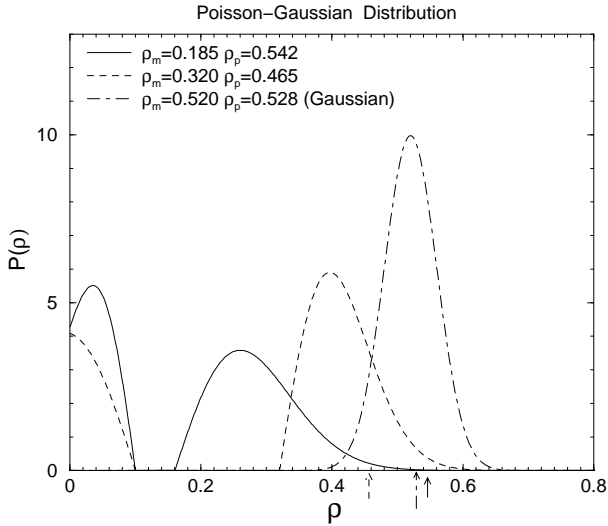}
\caption{Model charge distribution  for
the inhomogeneities or stripe two phase regions. The low density
insulating (antiferromagnetic) branch. The high
density hole-rich region starts at the compound average density $\rho_m$
and $\rho_p$, indicating by the arrows, 
is the density where percolation can occur.}
\end{figure}
An example of the distribution is shown in fig.1.
For illustration purpose, we show the results for compounds with
$\rho_m=0.185$ and $\sigma_+=0.05$
and with $\rho_m=0.32$ and $\sigma_+=0.038$. 
%The compound with 
%$\rho_m=0.18$ is in the overdoped region where the distribution  
%tends to be more  homogeneous. 
Above $\rho_m=0.25$ the charge distribution
becomes a simple Gaussian centered at $\rho_m$ with  $\sigma\le0.02$,
which reflects the non-existence
of the stripes phases and the observed homogeneous
charge distribution for the overdoped compounds.

The values of $\sigma$ for a given sample are chosen in order that
percolation in the hole-rich branch occurs exactly at
a given density $\rho_p$. Thus $T_c(\rho_p)$ is the maximum
temperature which the system can percolate and which we identify as
equal to $T_c(\rho_m)$.
Although we used a set of parameters to compare with the experimental
phase diagram of Bi2212, the main physical aspects can
be modeled by others distributions
with the similar results.
%This is the main point of this work and will be explained 
%in the end of the next section. 
According to percolation theory, percolation occurs in a square lattice
when 59\% of the sites or bonds are filled\cite{Stauffer}.
Thus, we find the density where the hole-rich branch percolates
integrating $\int P(\rho) d\rho$ from  $\rho_m$ till the integral
reaches the value of 0.59, where we define $\rho_p$. Below  $T_c(\rho_m)$
the system percolates and, consequently, 
it is able to hold a dissipationless supercurrent.
%if it is below a certain temperature which we will 
%identify as $T_c(\rho_m)$. 
To estimate $T_c(\rho_m)$
we need to calculate $T^*$ as function of $\rho$. 
%and this will be done in the next section.

\section{The Phase Diagram}
To develop the dynamics of the hole-type carriers in the Cu-O planes, we
adopt a two dimension extended Hubbard Hamiltonian 
in a square lattice of lattice parameter $a$
\begin{eqnarray}
H&=&-\sum_{\ll ij\gg \sigma}t_{ij}c_{i\sigma}^\dag
c_{j\sigma}+U\sum_{i}n_{i\uparrow}n_{i\downarrow} \nonumber \\
&& +\sum_{<ij>\sigma
\sigma^{\prime}}V_{ij}c_{i\sigma}^\dag c_{j\sigma^{\prime}}^\dag
c_{j\sigma^{\prime}}c_{i\sigma} 
\label{b}
\end{eqnarray}
where 
$t_{ij}$ is the hopping
integral between sites $i$ and $j$; $U$ is the Coulomb on-site correlated
repulsion and $V_{ij}$ is the a phenomenological attractive 
interaction between nearest-neighbor sites $i$ and $j$ which will
will argue later about its possible origin.
%which is, at this stage, a pure phenomenological potential. 

Using a BCS-type mean-field
approximation to develop Eq.(\ref{b}) in the momentum space, one
obtains the self-consistent gap equation, at finite 
temperatures~\cite{Mello96,Angi,Caixa}
\begin{eqnarray}
\Delta_{\bf k}=-\sum_{\bf k^{\prime}}V_{\bf
kk^{\prime}}\frac{\Delta_{\bf k^{\prime}}}{2E_{\bf
k^{\prime}}}\tanh\frac{E_{\bf k^{\prime}}}{2k_BT},%\label{cc}
\end{eqnarray}
with
$E_{\bf k}=\sqrt{\varepsilon_{\bf k}^2+\Delta_{\bf k}^2}, \label{rr}$
which contains the attractive potential $V_{\bf
kk^{\prime}}$ 
in the extended Hubbard Hamiltonian of Eq.2.
%Later we will argue, based on experimental evidences, 
%that it can be from polaronic origin. 
For  a d-wave order parameter, $U$ is summed out of the gap equation and
the amplitude of the attractive
potential $V$ is the only unknown variable and 
become an adjustable parameter\cite{Caixa,Caixa2}.
$\varepsilon_{\bf k}$ is a dispersion relation.
%Here we use parameters close to these, 
%used also before to derive the YBCO phase diagram\cite{Caixa2}.
The hole density and the gap equation, 
are solved self-consistently
for a d-wave order parameter\cite{Caixa,Caixa2}. 
%\begin{eqnarray}
%\rho(\mu,T)=\frac{1}{2}\sum_{\bf k}\left(
%{1-\frac{\varepsilon_{\bf k}}{E_{\bf k}} \tanh\frac{E_{\bf k}}
%{2k_BT}}\right) %\label{f}
%\end{eqnarray}
%where $0\leq \rho\leq 1$.

In Fig.2 we plot the temperatures of vanishing 
gap from Eq.2 which we take as $T^*$, as function of $\rho$.
Here we use  $V=-0.150$eV which reproduces well the experimental
values of $T^*\times \rho$ for the Bi2212 system\cite{Oda00}.
The  parameters are within 10\% to those taken from ARPES
measurements~\cite{Randeria} with a nearest
neighbor hopping  $t=0.12$eV and further hopping
parameters up to fifth neighbor. 
The particular form of the $T^*$ curve depends on the values of these
hoppings.
Notice that the density of holes  has a factor of 2
with respect to the values given by Ref.\cite{Oda00}, but in
agreement with ref.\cite{Angi} as it is more appropriate
for the Bi2212 system\cite{Konsin}.
%According the previous section, the values of $T_c$
%are found in the following manner: for a given density $\rho_m$, we
%calculate the value where percolation can occur, that is,  $\rho_p$,
%defined above.
%The $T^*$ associated with the density $\rho_p$ is the maximum
%percolating temperature, that is $T_c(\rho_m)\equiv T^*(\rho_p)\equiv T_c$.
\begin{figure}
\includegraphics[width=8cm]{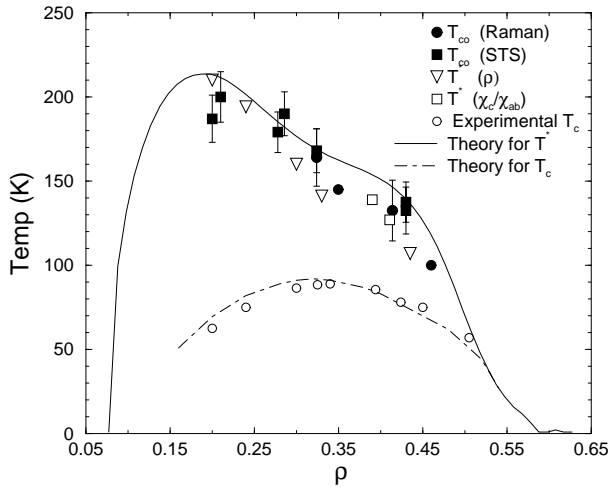}
\caption{The phase diagram taking $T^*(\rho)$ as the onset of vanishing
gap and $T_c(\rho_m)$ as the percolating threshold. The experimental
points and the symbols are taking from Ref.\cite{Oda00}}
\end{figure}

Thus the $T^*$ and $T_c$  shown in Fig.2, for a  sample with
average charge density $\rho_m$, are obtained in the
follow way: the  hole-rich  or metallic branch of the distribution  describes
the regions with hole charge densities $\rho \ge \rho_m$.
These charge fluctuations yield
clusters with local superconducting temperature $T_c(\rho)$, as 
$\rho$ varies in the sample and one may write $T_c(r)$, where
$r$ represents any position inside the compound.
For the metallic regions, $T_c(\rho)$ is a decreasing function of $\rho$ and
the maximum gap temperature occurs for $T^*(\rho_m)\equiv T^*$. The different
metallic regions in this sample have  $T_c(\rho) \le T^*$.
For temperatures below  $T^*$,
some superconducting clusters are formed, like small superconducting
islands embedding in a metallic and insulating medium.
Thus, as the temperature decreases, more clusters become
superconducting, and eventually the superconducting regions
percolates at $T_c$, that is, a superconducting current can flow for
temperatures $T \le T_c$.

\section{Discussion}

The fact that  $T^*$ decreases continuously with $\rho$
in the superconducting region ($\rho_m > 0.15$), as seen in Fig.2,
is very suggestive
and in agreement with the early ideas regarding a phonon
mediated superconducting interactions: Materials
whose vibrating atoms interact strongly with the electrons,
and are poor metals, should become superconductors at higher
temperatures than  those good metals, whose atoms interact
weakly with electrons\cite{Matthias}. 
For cuprates, as the doping level of
the samples increases, it is well known that the compounds change
from very poor metals in the
normal phase to  very good metals with typical Fermi liquid
behavior for overdoped samples. Since $T^* \times \rho$
is a decreasing curve, for any cuprate family, and assuming that
$T^*$ is  the onset of superconducting gap, such curve may be a
strong indication of the phononic superconducting interaction.

There are several observations and
measurements that can be well explained within the percolating approach,
we will discuss here only a few examples.

%1-ARPES measurement\cite{Harris} of the zero temperature gap $\Delta_0$ 
%have shown that, differently from the usual superconductors, it is 
%uncorrelated with $T_c$. 
1-Harris et al\cite{Harris}, through ARPES measurements, 
have reported the anomalous behavior of $\Delta_0(\rho_m)$ 
which decreases steadily with the
doping $\rho_m$ although $T_c$ increases by a factor of 2 for 
their underdoped samples.
In the overdoped region, since $T_c$ also decreases, the behavior
is the expected conventional proportionality.
It is well known that superconductors
have a constant value for the ratio $2\Delta_0/k_BT_c$, being
3.75 for usual isotropic order parameter and 4.18 for 
$d_{x^2-y^2}$ solution\cite{Maki}.       

\begin{figure}
\includegraphics[width=8cm]{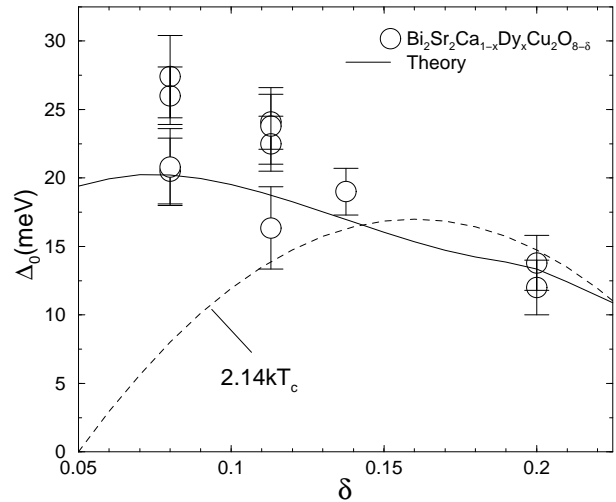}
\caption{The zero temperature gap for 9 samples as measured by Harris et
al\cite{Harris} and  our calculations.}
\end{figure}
%As we pointed out above, the large inhomogeneities
%in the underdoped region creates regions with different
%superconducting temperature and the critical temperature
%Since $T_c$ is a superconducting percolation threshold which is consequence of
%the charge distribution and it is completely uncorrelated
%with  $\Delta_0$. 
At low temperature, since the superconducting region 
percolates through different regions, each with a given $\Delta_0(r)$, 
tunneling and ARPES
experiments detect the largest gap present in the compound. 
Consequently, $\Delta_0(\rho_m)$ must be correlated with the onset
of vanishing gap $T^*(\rho_m)$ which is the largest superconducting
temperature in the sample, and not with $T_c(\rho_m)$. As we show in
Fig.3, correlating the values plotted in Fig.2 for $T^*(\rho_m)$
with $\Delta_0(\rho_m)$, we are able to give a reasonable 
fit for the data of Harris et al\cite{Harris} on $Dy-BSCCO$
and explains the different energy scales pointed out by
Harris et al and several others authors.

2- The resistivity measurement is also one of the tools to detect the pseudogap.
The underdoped and optimum doped high-$T_c$ oxides have a linear
behavior for the resistivity  in the normal phase 
up to very high temperature.
However, at $T^*$ there is a deviation from the linear behavior and
the resistivity falls faster with decreasing temperature\cite{TS,Oda00}.
This behavior can be understood by our model,
with the increasing of superconducting cluster numbers and size, as the
temperatures decreases below $T^*$. Each superconducting cluster produces
a short circuit which decreases the resistivity below the
linear behavior between $T^*$ and  $T_c$. 

3- Recently measurements of magnetic domains  above $T_c$
has been interpreted as a diamagnetic precursor to the
Meissner state, produced by performed pairs in underdoped
$La_{2-x}Sr_xCuO_4$ thin films\cite{IYS}. The existence of
superconducting cluster between $T^*$ and  $T_c$ easily
explains the appearance of local diamagnetic or Meissner domains, and,
if there is a temperature gradient in the sample, the
local flux flows and produces the dynamic flux flow state\cite{Xu}.

4- Another important consequence is that 
the pairing mechanism must be investigated by experiments
performed mainly at $T^*$.
A such experiment was accomplished by Rubio 
Temprano et al\cite{Temprano}, which measured a large isotope
effect  associated with $T^*$  and an almost 
negligible isotopic effect associated with $T_c$ 
in the slightly underdoped $HoBa_2Cu_4O_8$ compound.
The results strongly support the fact that 
electron-phonon induced effects are present in the superconducting
mechanism associated with $T^*$ and  with the percolation
approach to $T_c$. 
%%Bussmann-Holder et al have also calculated
%%$T^*$ as function of a phonon induced gap\cite{Annette}.
%In order to gain further insight on
%the nature of the pair potential,
%we and our colleagues have measured  the resistivity under
%hydrostatic pressure on an
%optimally doped $Hg_{0.82}Re_{0.18}Ba_2Ca_2Cu_3O_{8+\delta}$\cite{futurework}
%sample.  The  data
%indicate a linear increase of $T^*$ with the pressure. In the
%context of our theory, this result also
%supports the phonon induced mechanism: the inhomogeneities or different
%local charge densities in a given compound yield different values
%for the Fermi level, broadening $N(E_F)$\cite{OWK}. The applied pressure
%on a cuprate with an inhomogeneous charge distribution
%is also expected to broaden the density of states\cite{Angi} 
%$N(E_F)$ and
%the main effect of an applied pressure on  $T_c$
%is an  increase of the phonon frequency or Debye frequency.
%This is seen through the linear
%increase on $T^*$\cite{futurework} which also provides, for the first time,
%a physical explanation on the origin of the linear pressure induced
%{\it intrinsic effect}\cite{Angi,Orlando,Jorge} usually postulated to explain
%the raise of $T_c$ above its maximum value.
%To obtain
%a quantitative fitting of this effect, we are presently working 
%on a simulation for the resistivity of a linear metallic medium
%with short circuited regions which varies with the temperature.
%
\section{Conclusions}
%
%We have shown that it is possible to describe
%several properties for high-$T_c$ oxides by a method which 
%takes into account the well documented charge inhomogeneities. 
We have demonstrated that  the percolating approach for an
inhomogeneous charge distribution  on the $CuO_2$ planes provides 
new physical explanations 
for many experiments performed on high-$T_c$ oxides
and quantitative results for their phase diagram.
%The inhomogeneous charge distribution
%is described phenomenologically in terms of a distribution
%which mimics the stripes phases. 
%Our novel method provides new insights
%on the physics of these superconductors. 
Contrary to some current
trends, which $T_c$ is regarded as a phase
coherence temperature and the existence of a gap phase without coherence
between  $T_c$ and $T^*$, in our approach,  $T_c$ is a percolating
temperature for different regions which, due to the inhomogeneities,
possess different local superconducting transition $T^*(r)$. 
Similarly, instead of having a superconducting
gap $\Delta_{sc}$ and an excitation gap $\Delta$ associated
with $T^*$, we have a distribution of locally dependent $\Delta_{sc}(r)$. 

The  method described in this work can be applied in any cuprate and
yields also several new implications which will be
discussed elsewhere, but one of the most
interesting is that one can search for materials with very large $T_c$'s if
a better control of the local doping level is achieved.

Financial support of CNPq and FAPERJ is gratefully acknowledged.
JLG thanks CLAF for a CLAF/CNPq pos-doctoral fellowship.

\end{multicols}

\begin{references} 
\bibitem{BM} J.G. Bednorz and K.A. M\"uller, Z. Phys., {\bf 64}, 189 (1986).
%
\bibitem{TS} T. Timusk and B. Statt, Rep. Prog. Phys., {\bf 62}, 61 (1999).
%
\bibitem{Retal98} C. Renner, B. Revaz, J.-Y Genoud, K. Kadowaki,                 and O. Fischer, Phys. Rev. Lett. {\bf 80}, 149 (1998).
%
\bibitem{Setal99} M. Suzuki, T. Watanabe, A. Matsuda, Phys. 
Rev. Lett. {\bf 82}, 5365 (1999).
%
\bibitem{Shen95} Z-X. Shen and D. S. Dassau, Phys. Rep., {\bf 253}, 1 (1995).
%
\bibitem{Ding96} H. Ding, T. Yokkoya, J.C. Campuzano, T. Takahashi,
M. Randeria, M.R Norman, T. Mochiku, K. Kadowaki, and J. Giapintzaki, 
Nature, {\bf 382}, 51 {1996}
%
\bibitem{Oda00} M. Oda, N.  Momono and M. Ido, Supercond. Sci. Technol. {\bf
13}, R139, (2000).
%
\bibitem{Takagi92} H. Takagi, B. Batlogg, H.L. Kao, R.J. Cava, 
J.J. Krajewski, and W.F. Peck,  Phys. Rev. Lett. {\bf 62}, 2975 (1992).
%
\bibitem{Loram97} J. Loram, K.A. Mirza, J.R. Cooper, J.L. Tallon, 
Physica {\bf C282-287}, 1405 (1997).
%
\bibitem{Billinge00} S.J.L. Billinge, et al, J. Supercond., Proceedings of the
Conf. "Major Trends in Superconductivity in New Milenium", 2000.
%
\bibitem{Buzin00} E.S. Bozin, G.H. Kwei, H. Takagi, and S.J.L. Billinge, 
Phys. Rev. Lett. {\bf 84}, 5856, (2000).
%
\bibitem{Pan} S.H.Pan et al, cond-mat/0107347.
%
\bibitem{Egami96} T. Egami and S.J.L. Billinge, in "Physical Properties
of High-Temperatures Superconductors V" edited by D.M. Greensberg,
World Scientific, Singapure 1996), p. 265.
%
\bibitem{Tranquada} J.M.Traquada,  B.J. Sternlieb, J.D, Axe, Y. Nakamura,        and S. Uchida, Nature (London),{\bf 375}, 561 (1995).
%
%\bibitem{Bianconi} A. Bianconi et al, Phys. Rev. Lett. {\bf 76}, 3412 (1996).
%
%\bibitem{EKT} V.J. Emery, S.A. Kivelson and J.M.Traquada, Proc. Natl.
%Acad. Sci. USA {\bf 96}, 8814 (1999).
%
\bibitem{Xu} Z.A. Xu, N.P. Ong, Y. Wang, T. Kakeshita, and S. Uchida,
Nature {\bf 406}, 486-488 (2000). 
%
\bibitem{IYS} I. Iguchi, I. Yamaguchi, and A. Sugimoto, 
Nature, {\bf 412}, 420 (2001).
%
\bibitem{OWK} Yu.N. Ovchinnikov, S.A. Wolf, V.Z. Krezin, Phys. Rev. {\bf B63},
6452, (2001), and Physica {\bf C341-348}, 103, (2000).
%
\bibitem{Egami} T. Egami, Proc. of the New3SC International Conference,
to be published in Physica C.
%
\bibitem{Stauffer} D.F. Stauffer and A. Aharony, "Introduction to 
Percolation Theory". Taylor\&Francis Ed., London, 1994.
%
\bibitem{Mello96}
E.V.L. de Mello, Physica {\bf C259}, 109 (1996).
%
\bibitem{Angi}
G.G.N. Angilella, R. Pucci, and F. Siringo, Phys. Rev. B {\bf
54}, 15471 (1996).
%
\bibitem{Caixa}
E.S. Caixeiro, and E.V.L. de Mello, Physica {\bf C353}, 103 (2001).
%
\bibitem{Caixa2}
E. S. Caixeiro, and E.V.L. de Mello, submitted to the J. Phys. {\bf CM}.
%
\bibitem{Randeria} M.R. Norman, M. Randeria, H. Ding, J.C. Campuzano,
and A.F Bellman, Phys. Rev. {\bf B52}, 615 (1995).
%
\bibitem{Konsin} P. Konsin,  N. Kristoffel, and B. Sorkin, 
J. Phys. C.M. {\bf 10}, 6533 (1998).
%
\bibitem{Matthias} B.T. Matthias, "Superconductivity", Scientific
American, 92, November of 1957.
%
%\bibitem{Schabel}
%M.C. Schabel, C.-H. Park, A. Matsuura, Z.-X. Shen, Phys. Rev. B {\bf 57},
%6090 (1998).
%
%\bibitem{Oda} M. Oda, N. Momono and M. Ido, Supercond. Sc. Technol
%{\bf 13}, R139, (2000).
%
\bibitem{Harris} J.M. Harris, Z.H. Shen, P.J. White, D.S. Marshall,             M.C. Schabel, J.N. Eckstein, and I. Bozovic, Phys.Rev. B{\bf 54} R15665 (1996).
%
\bibitem{Maki} H. Won and K. Maki, Phys. Rev. {\bf B49}, 1397 (1994).
%:
\bibitem{Temprano}D. Rubio Temprano, J. Mesot, S. Janssen, K. Conder,           A. Furrer, H. Mutka, and K.A. M\"uller, Phys.Rev.Lett {\bf 84}, 1990 (2000).
%
%\bibitem{Annette} A. Bussmann-Holder et al, J.Phys. C.M. {\bf 13}, L169, 2001.
%
%\bibitem{futurework} E.V.L. de Mello et al, to be published.
%
\bibitem{Orlando} M.T.D. Orlando, A.G Cunha,  E.V.L. de Mello,                  H.Belich,  E. Baggio-Saitovich, A. Sin, X. Obradors, T. Burghardt,              A. Eichler,  Phys. Rev. {\bf B61}, 15454 (2000).
%
\bibitem{Jorge} J.L. Gonzal\'ez, M.T.D. Orlando, E.S. Yugue, E.V.L. de Mello
and E. Baggio-Saitovich, Phys. Rev. {\bf B63}, 54516 (2001).
%

\end{references}
\end{document}